# Chapter XX

# Technosocial risks of 'ideal' emotion recognition technologies: A defense of the (social) value of emotional expressions

Alexandra Prégent[1]

**Abstract**

The prospect of AI systems that I call "ideal" emotion recognition technologies (ERTs) is often defended on the assumption that social life would benefit from increased affective transparency. This paper challenges that assumption by examining the technosocial risks posed by ideal ERTs, understood as multimodal systems capable of reliably inferring inner affective states in real time. Drawing on philosophical accounts of emotional expression and social practice, as well as empirical work in affective science and social psychology, I argue that the appeal of such systems rests on a misunderstanding of the social functions of emotional expression. Emotional expressions function not only as read-outs of inner states, but also as tools for coordinating action, enabling moral repair, sustaining interpersonal trust, and supporting collective norms. These functions depend on a background of partial opacity and epistemic friction. When deployed in socially authoritative or evaluative contexts, ideal ERTs threaten this expressive space by collapsing epistemic friction, displacing relational meaning with technology-mediated affective profiles, and narrowing the space for aspirational and role-sensitive expressions. The result is a drift towards affective determinism and ambient forms of affective auditing, which undermine both social cohesion and individual agency. I argue that, although it is intuitive to think that increasing accuracy would legitimise such systems, in the case of ERTs accuracy does not straightforwardly justify their deployment, and may, in some contexts, provide a reason for regulatory restraint. I conclude by defending a function-first regulatory approach that treats expressive discretion and intentional emotional expression as constitutive of certain social goods, and that accordingly seeks to protect these goods from excessive affective legibility.

# 1 Introduction

The prospect of new technological threats often reveals the hidden value of certain things. As emotion recognition technologies (ERTs) are increasingly capable of inferring affective states, the prospect of future, fully functional ERTs, understood as multimodal AI systems capable of reliably inferring inner affective states in real time, has moved from science fiction to a real technosocial possibility. Ethical and legal debate has so far (rightly) concentrated its attention on issues surrounding bias, technical limitations, and scientific validity, which are the main concerns regarding currently deployed systems (Barrett et al. 2019; Crawford et al. 2019). But in so doing, while ERTs keep improving, the ethical discourse around them has yet to formulate principled reasons for constraining future systems that actually "work as intended". In the remainder of this chapter, I will refer to such systems as "ideal ERTs". Hence, when confronted with the normative conundrum of whether or not society would benefit from ERTs rendering individuals' inner affective life legible, many proponents of their use, more often than not the companies developing them, argue that this technology brings better transparency and honesty vis-à-vis one's own inner affective life and that of others (see iMotions, SmartEye, MorphCast, Kopernica). In contrast, privacy advocates typically point to the (early liberal) idea in privacy theory that posits that individuals need time away from society and constant scrutiny to develop their own agency and flourish (Menges & Weber-Guskar 2025; Akar 2024; Ventham 2025).

---

[1] A. Prégent (Corresponding Author)
  Philosophy Department, Leiden University, Leiden, The Netherlands
  e-mail: a.pregent@phil.leidenuniv.nl

This chapter proposes to look at the *social reasons* behind the technosocial risks that the use of ERTs poses. The argument advanced in this chapter is *not* that individuals would *not* benefit from such a protected space, but rather that social reasons should be *added* as another type of harm. In that sense, I do not propose a competing argument but a complementary one, which, in my view, strengthens the privacy claim in favour of ERT regulation.

Specifically, the chapter argues that by rendering inner affective life legible, AI systems that successfully infer people's emotional states (ERTs) impinge on core social functions of (intentional) emotional expressions, functions that, it argues, matter for the well-being of our collective social life. Drawing on empirical research in affective science and social psychology, as well as philosophical work on expression, signalling, and social practices, I discuss how emotional expressions function not merely as indicators of inner states, but as tools for coordinating action, facilitating moral repair, shaping interpersonal trust, sustaining collective emotions, and enabling normative flexibility. Through these functions, I argue, emotional expressions constitute a socially distributed normative infrastructure that produces significant interpersonal, epistemic, moral, and political goods. I then introduce "ideal" ERTs and show how their deployment threatens to undermine the very goods emotional expressions produce.

The structure of the argument is as follows. Section 2 conceives intentional emotional expressions as a social practice and analyses the kind of goods they generate, notably through particular attention to their epistemic, moral-relational, and collective-political functions. Section 3 introduces the thought experiment of ideal ERTs and examines how their deployment would transform the ecology within which intentional emotional expressions operate. Section 4 articulates the kinds of harm that result, framed as losses of social goods, and draws out the implications for regulation, arguing that intentional emotional expressions should be treated as an overarching public good that requires protection from excessive affective legibility.

## 2 Intentional emotional expressions as a social practice

Understanding the technosocial risks of ideal ERTs requires a prior account of the social value of intentional emotional expressions. In ordinary life, emotional expressions (EEs) play a distinct social-epistemic role. They provide partial and defeasible information about how others evaluate a situation, what they care about, and how they are likely to respond. In that sense, they act as socially embedded tools that regulate interactions, support cooperative action, help maintain interpersonal relationships, and contribute to the stability of social norms. Their functions are thus rich and multidimensional, and they play a central role in shaping the patterns of behaviour that structure collective life.

A large body of work in social and affective psychology has converged on the idea that emotional expressions are not mere epiphenomena of inner feelings, but also function as socially embedded signals that influence how others think, feel, and act (van Kleef 2016; Hess & Fischer 2013; Scarantino 2017, 2019). On this view, emotional expressions help to shape the interpersonal environment through which individuals constantly navigate. Van Kleef's Emotions as Social Information (EASI) model, for instance, shows that observers routinely draw inferences about others' appraisals, goals, and intentions from their emotional expressions, and that these inferences systematically guide behaviour in negotiation, group interaction, and decision-making (van Kleef 2009, 2016).

Scarantino's theory of affective pragmatics further posits that emotional expressions are a type of communicative move that can be used to express what is inside, to represent the world, to direct other people's behaviour, and to commit to future courses of action, in analogy to what we do with speech acts (Scarantino 2017). An angry outburst can signal that a norm has been violated. A fearful expression can direct others' attention to a possible threat. A warm smile can signal acceptance and willingness to cooperate. In each case, the expression is part of a normative practice that conveys social information and shapes interaction. Affective researchers typically distinguish between two types of EEs: spontaneous EEs and intentional EEs. Spontaneous EEs are generally considered to be involuntarily produced, while intentional EEs are voluntarily produced by the emoter. This dichotomy is not as clean-cut as it may seem,[2] and prototypical EEs (e.g. smiling to express joy, frowning to express anger) may often be used in both cases. For example, a person may smile spontaneously when feeling happy, but may also smile intentionally (e.g. when a mother encourages her toddler to take a few more steps). Spontaneous EEs often track more accurately inner affective states and are thus highly favoured by affective computing researchers (Zhang et al. 2020) This is so because they are not voluntarily produced, contrary to intentional EEs. However, spontaneous EEs do little social and normative work in our everyday life, in contrast to intentional EEs.

*Intentional* emotional expressions are often produced against the background of display rules and social norms.

---

[2] Within this dichotomy are many "edge cases" such as semi-natural EEs, ad hoc EEs, idiosyncratic EEs, or even semi-conventional EEs (see Bar-On 2004, 2010). Some accounts will, for instance, reject the possibility that EEs may be insincere or intentional by distinguishing signals from cues and counting the latter as the only *genuine* EE for instance (see Green 2007).

While these rules differ across cultures, roles, and relationships, people nevertheless learn when, where, and to whom different emotional expressions are appropriate and, given, for instance, the context, how they should be modulated in intensity (Matsumoto, Yoo, & Fontaine 2008). The same expression of anger that is acceptable in a political protest may be inappropriate in a formal meeting. This is because the meaning of the expression shifts through, and is partly inferred from, the specific context, but also from the presence or absence of different perceivers, as well as from any other situational factors that might influence the meaning (Fischer & Sauter 2017). Because observers cannot fully access the inner mental states of others, they must interpret expressions in light of incomplete evidence. This affective opacity is often valuable. A polite smile that masks mild irritation can preserve the relationship and leave room for future negotiation. A controlled expression of disapproval can create space for the other to correct their course without facing humiliation.

Yet this opacity is also not so high that perceivers cannot successfully interpret the meaning of EEs in a reliable way either. In fact, many moral practices rely on the possibility of expressing and perceiving emotional stances in ways that are intelligible to others. Apologies, expressions of gratitude, forgiveness, reassurance, encouragement, admiration, and sympathy are all bound up with emotional expression. Philosophers of moral psychology have long emphasised the importance of recognitional attitudes such as respect, resentment, gratitude, and indignation for moral life (Strawson 1962; Wallace 1994). Empirical work strengthens these insights by further showing that these attitudes are often communicated and perceived through patterns of expression in the face, voice, and posture (Hess and Hareli 2012; Parkinson et al. 2005).

Consider the cases of apology and forgiveness. An apology that is delivered with a flat tone and no visible signs of contrition is typically judged to be insincere. The same words, accompanied by a lowered gaze, a softened voice, and visible regret, are usually judged quite differently. Recipients pay attention to EEs and treat them as indicators of acknowledgement and commitment to change (see Smith 2008). In turn, forgiveness often involves expressive signals that communicate the release of resentment and a willingness to restore or transform the relationship. These expressive dynamics are central to moral repair (see Walker 2006). They allow parties to signal changing attitudes and to coordinate new expectations.[3]

Expressions are equally important for communicative acts of care. Reassurance requires expressions of warmth and concern in order to be effective, especially in contexts of vulnerability such as illness or grief. Sympathy requires expressions that acknowledge the other's suffering. Gratitude requires expressions that highlight the significance of another's action (Weiss 1985; Schwartz 2020; cf. Searle 1979). These expressions do not merely accompany the underlying attitudes but are part of what it is to enact these attitudes in a shared social space.

What this demonstrates is that the relational goods that arise from expressive practices depend on more than direct access to inner states. They depend on a capacity for modulation and selective disclosure that is responsive to context and to the vulnerability of others (Goffman 1959; Baier 1986). People routinely temper their expressions to protect the feelings of those they care about, to avoid unnecessary conflict, or to maintain dignity (Goffman 1959). For instance, a teacher may suppress visible anger in order not to shame a student. A friend may soften expressions of disappointment in order to preserve a relationship that is otherwise supportive. These modulations are not necessarily deceptive and are best understood as expressions of respect for others in an ongoing interaction or relationship.

The social-epistemic functions of intentional EEs are thus characterised by two distinctive features. First, they are context-sensitive. Observers interpret expressions by integrating them with background knowledge about the situation and the person expressing them. Second, they are non-deterministic. Observers treat expressions as evidence that can be revised and refined rather than as conclusive proof (Scarantino 2017; Fischer & Sauter 2017). This introduces what we might call epistemic friction. The presence of friction encourages agents to remain open to correction and to seek clarification. These features of expressive practice are central to social cohesion (van Kleef 2016) and, importantly, *depend* on the fact that expressions *do not* offer unmediated access to inner affective life.

The ability to modulate and negotiate one's EEs in a given context according to role and relationship is part of what allows people to fulfil complex moral and relational expectations. When expressions are subject to external classification that focuses on exposing the inner affective life, the space for such modulation is narrowed. The relational meaning of the expression is displaced by the system's judgement about its "true" emotional content.

Ideal ERTs thus threaten the social-epistemic functions of EEs by replacing defeasible, context-sensitive interpretation with diagnostic classifications.

## 3 The threat of ideal ERTs

---

[3] Strawson (1962) for instance, states that "To ask to be forgiven is in part to acknowledge that the attitude displayed in our actions was such as might properly be resented and in part to repudiate that attitude for the future (or at least for the immediate future); and to forgive is to accept the repudiation and to forswear the resentment." (p.191).

To isolate the structural risks posed by ideal ERTs, it is useful to set aside, for the moment, familiar concerns about their present scientific shortcomings. Several influential reviews have documented that many current emotion recognition systems rely on oversimplified emotion models and that their accuracy and robustness are often limited, especially in real-world conditions (Barrett et al. 2019; Crawford 2021; Wang et al. 2022). The prospect of ideal ERTs requires us to depart from these criticisms. The question guiding the present chapter concerns *what happens when affective computing systems become sufficiently accurate to be widely trusted for practical purposes*.

Recent surveys indicate that multimodal affective computing, which typically combines facial, vocal, behavioural, and physiological signals, has achieved significant performance gains in controlled conditions and is increasingly being deployed in commercial applications spanning healthcare, workplaces, education systems, and law enforcement (McStay 2018; Wang et al. 2022). It is therefore reasonable, for the sake of normative analysis, to consider a future in which idealised versions of these systems are embedded in many aspects of social life.

An ideal ERT, in the sense used here, is a system that achieves high accuracy in recognising a broad range of emotional expressions from (presumably multimodal) inputs. It can operate in real time and can adjust for both individual and cultural variation (a presumption of adaptivity). The reader might feel like I have under-described these systems and that their definition is thus rather vague. While I aim to be as precise as possible, my goal is not to forecast how, technically, these ideal ERTs will come to see the light of day, but rather to focus instead on the technosocial risks that their development would engender.

The introduction of ideal ERTs risks generating several deep social transformations in how EEs function.[4] They do so by introducing a new layer of interpretation between emoter and perceiver. With ideal ERTs, instead of observers forming their own (defeasible) inferences about what an EE signals, they are now presented with output data that purport to represent inner emotional states (and other related information such as intensity, co-occurrence with other emotional states, etc.).

From a social-epistemic standpoint, this supplements perceivers' standard recognition process and potentially replaces it with a different mode of epistemic access. The 'ordinary', human-based mode is characterised by fallibility, context-sensitivity, and interpersonal negotiation. It encompasses the meaning conveyed by the full range of emotional expressions, from spontaneous to intentional EEs. The 'new', technologically based mode is characterised by diagnostic outputs. It focuses on revealing inner emotional states, disregarding EEs' social meaning. This shift in interpretive practices has two social-epistemic effects. First, it undermines the epistemic friction that is essential for some expressive meaning and thereby weakens the very processes that make intentional EEs reliable sources of social information. Second, it encourages a normative focus on the correspondence between inner affective states and outward emotional expressions, rather than on the relational and practice-dependent functions of expressive acts.

Regarding the first effect, by reducing the epistemic friction in affective communication, ERTs impinge on the functions of intentional EEs, which depend on this friction to successfully communicate their meaning. As explained in Section 2, intentional EEs are valuable not because they give direct access to inner states, but because they provide signals that are interpretable enough to guide social interaction, yet opaque enough to enable the emoter to have reasonable control over what type of social meaning their EEs will convey.

Consider the example of apology. In ordinary interpersonal life, an apologiser might intentionally modulate their tone, gaze, or posture in order to convey contrition. This expression might not necessarily mirror the exact emotional state they feel. Instead, it enacts a recognitional attitude that communicates acknowledgement of wrongdoing and willingness to repair (Walker 2006; Hess and Hareli 2015; Parkinson et al. 2005). The success of apology depends on the recipient treating these expressions as *sincere* within the context of the relationship, not on the expressions matching some inner affective guilt. Sometimes we may still feel irritation even though deep down we know that we want to apologise and that we are determined to work on moving beyond this emotion of, let's say, "irritation". Under conditions of real-time ERT monitoring, however, this absence of genuine guilt is instantly revealed to the recipient. If the system labels the person's underlying affect as "irritation", "anger", or even "frustration", the carefully modulated apology may fail, not because it was insincere, but because the technological mediation displaces the relational meaning of the intentional expression. The very possibility of moral repair is thus under threat.

A similar dynamic plays out in reassurance. A parent or clinician may suppress worry or fear in order to comfort a child or patient. A calm tone and warm expression can communicate safety even when the underlying emotion is mixed. This form of modulation is not a form of deception but part of what it means to provide care in a vulnerable situation. Yet ideal ERTs, by surfacing the underlying affective profile in real time, may strip reassurance of its pragmatic force. If the system reveals that the parent's or clinician's heart rate, (micro) facial tension, or vocal stress correlates with fear or anxiety, *the intentional expression loses its meaning as an act of care*.

---

[4] On the idea of deep disruption, see Hopster (2021).

Politeness routines are another example of communicative acts that rely on emotional expression to succeed and therefore depend on partial opacity. In many contexts, individuals intentionally use polite smiles or neutral expressions to maintain harmony, preserve face, and prevent unnecessary conflict. These expressions function as social lubricants that allow cooperation despite minor irritations or differences in temperament (Fischer & Sauter 2017). Ideal ERTs, by highlighting the presence of anger or irritation beneath the polite surface, remove this space for successful social navigation. A polite smile that conceals mild frustration is reclassified as insincere, and the social work it performs is undermined. Instead of reducing tension, the technologically revealed affect can thus escalate it.

The same problem arises in educational contexts. A teacher might intentionally hide frustration in order not to discourage a struggling student. A soft tone or neutral expression helps maintain the student's confidence. Yet if ideal ERTs detect and display the teacher's underlying frustration to supervisors or students, the teacher's pedagogical strategy becomes ineffective. The expression is no longer taken at face value but is filtered through the system's inference of "frustration". The relational meaning of the expression is overridden, and the teaching relationship is altered in ways that neither participant necessarily intends.

Across all these cases, the epistemic function of intentional EEs is not to reveal inner states but to shape the relational context.[5] Their value lies in how they guide social behaviour, not in how accurately they map inner affective states. The partial opacity between expression and underlying affect gives individuals the space to act as good friends, good teachers, good colleagues, and good caregivers, even when their emotions are mixed, ambivalent, or in tension with their normative aims.

Ideal ERTs disrupt this social-epistemic structure. As a result, intentional EEs cease to function as effective signals and the very capacities that intentional EEs enable, namely coordination and general social cohesion, are jeopardised.

This is a transformation of the social-epistemic ecology within which expressive practices operate. The normative functions of intentional EEs depend on a shared understanding that emotional expressions are communicative tools whose social meaning is not reducible to inner affective states. When ideal ERTs treat the latter as the true content of the expression, the communicative space in which intentional EEs operate collapses. This shift endangers the very goods that intentional EEs provide in ordinary interaction and, if widely deployed, risks reshaping social norms around a conception of affective transparency that is deeply at odds with the requirements of social cohesion.

Whereas the first effect concerns the collapse of epistemic friction by revealing inner affective states in real time, the second concerns the structural transformation of how emotional expressions are interpreted in social life. Intentional EEs acquire their meaning through a complex interplay of situational factors, interpersonal history, social roles, and normative expectations. The same expression of anger can signify moral protest, playful teasing, self-directed frustration, or boundary setting depending on the context. The same expression of sadness can represent grief, disappointment, reflection, or an attempt to solicit support. As Fischer and Sauter emphasise, emotional expressions are deeply context-dependent and cannot be properly understood without reference to the social norms and relational dynamics in which they are embedded (Fischer & Sauter 2017; see also (Colombetti & Krueger 2015; Barrett 2017; Barrett et al. 2019). Human interpreters routinely integrate these factors into their judgements. They recognise, for example, that a sarcastic remark delivered with exaggerated irritation has a different meaning in a close friendship than in a strained professional environment. Ideal ERTs disrupt this interpretive ecology by abstracting expressions away from their local contexts. They will (most likely) treat emotional expressions as data signatures that map onto (discrete) emotion labels rather than as acts whose meaning emerges through social interaction. This brings the focus not on the EE's social function, but on the coherence between the ERTs' inner state-related output and the EE, bringing the relational meaning into second place. Observers are invited to think that what really matters is *whether an expression transparently discloses all relevant affective components*, rather than whether the expression performs its relational role well.

One might object that social practices are historically plastic, and that norms surrounding emotional expression could adapt to conditions of affective transparency, much as norms of privacy and disclosure have adapted to other technological changes.[6] To be clear, what I am arguing here is not merely that partial opacity and epistemic friction are features of our social practices, but that they are in fact *constitutive conditions* for certain social goods, especially in the cases of moral repair, aspirational commitment, role-based trust, and social coordination. Hence, the argument is not that existing practices would be slightly disrupted while adapting to the new reality that ideal ERTs would bring about, but rather that the goods those practices realise depend on epistemic features, namely defeasibility, negotiability, and expressive discretion, that *cannot function under conditions of systematic affective transparency*. Certain social goods depend on agents' ability to act and be responded to *without* their expressive acts being continuously overridden by authoritative interpretations of their inner affective states. Moral repair, aspirational commitment, role-based trust, and social coordination require that intentional emotional expressions retain a degree of normative authority independent of their alignment with inner affective states.

---

[5] On the related topic of affective scaffolding, see Colombetti & Krueger (2015).

[6] I want to thank a reviewer for raising this objection.

While alternative practices might emerge under such conditions, they would not be functionally equivalent. A practice in which forgiveness, reassurance, or apology is normatively assessed by reference to detected affective states does not preserve the same social good as one in which these expressions serve to initiate and stabilise commitments amid affective ambivalence. What is lost is not merely a familiar way of interacting, but the possibility of taking responsibility for a stance before one's affective life has fully aligned with it. This loss cannot be compensated by new norms of transparency without transforming the underlying conception of agency that these practices presuppose. But aspirational expressions are only one instance of the broader phenomenon I am pointing to. Similar losses arise in other practices such as moral repair, role-based trust, and social coordination, all of which depend on the capacity to express a stance whose normative force is not overridden by contemporaneous affective states. In each case, intentional emotional expressions function as acts of responsibility, respect, or reliance rather than as reports of inner affect. Systematic affective transparency does not merely alter how these practices are interpreted; it undermines the conditions under which they can realise the same social goods at all.

## 3.1 Emotion recognition: disentangling human from machine

I have so far kept an important distinction implicit.[7] It is now time to address it directly in order to clarify how one can plausibly maintain, at once, that ideal ERTs may infer emotional states in the near future and that emotional expressions are complex, context-dependent forms of communication.

The apparent tension arises if one moves too quickly from claims about emotional expression to claims about emotional recognition. At first sight, the possibility conditions of ideal ERTs may appear to conflict with the way emotions and emotional expressions have been characterised here: as context-sensitive, potentially plural, and in constant fluctuation. This tension seems especially acute in light of constructionist theories of emotion (Barrett 2017).

Before attempting to dissolve it, recall that ideal ERTs, as defined here, are not extensions of current affect-detection approaches. Existing systems typically rely on the assumption that simple, prototypical expressions (such as smiling, frowning, or pupil dilation) transparently track discrete inner emotional states across individuals, cultures, genders, and ages. This simplified view maps to the (simplified) basic emotion theory, which holds that a limited set of biologically innate emotions constitute natural kinds, each associated with a distinctive and relatively invariant expressive and physiological signature that allows for reliable recognition across contexts. In contrast, constructionists typically hold that emotions are constructed psychological events that emerge from more basic affective and cognitive processes in context.

The prospect of ideal ERTs is to be positioned between the simplified versions of basic emotion theory that dominate much affective computing (e.g. Ekman 1994) and constructionist accounts of emotion (e.g. Barrett 2017).

Positioned in between these two theories of emotion, finely tuned, multimodal, and adaptive ERT systems could reliably infer affective states through various metrics expanding classic emotion-type (anger, sadness, or happiness) by integrating other evaluative models such as valence, arousal, and dominance models (Yoon 2022). Emotion-types, understood as coarse-grained categories used for practical purposes, would then be used only for simplifying the outputs,[8] not the primary basis of the inference model. Ideal ERTs, as understood here, do not presuppose nor need to adopt a strong essentialist account of emotions as being distinct, unique, and clear-cut categories (i.e. emotion kinds). They require only that inner affective states, however contextually shaped and multiply realised, exhibit sufficient regularity at a basic categorical level to be inferred with high reliability. On this understanding, accepting Barrett's critique of basic emotion theory (2017, 2019) is compatible with granting, for the sake of the present argument, that ideal ERTs could accurately detect inner affective states in a limited yet socially significant sense. The apparent contradiction is therefore only prima facie.

Barrett's critique (2017, 2019) of current ERTs, for instance, is directed at strong claims about emotions as *biologically fixed natural kinds* with *context-independent signature*s. But this critique of the grounds (or lack thereof) on which the basic emotion theory relies does not, however, exclude the possibility that finely tuned, multimodal, and adaptive systems could reliably infer affective states (such as anger, sadness, or happiness) understood as *simplified categories used for practical purposes but nonetheless relying on complex and context-sensitive patterns of inference*. Ideal ERTs, as understood here, therefore do not presuppose nor need to adopt a strong essentialist account of emotions as being distinct and clear-cut categories. They require only that inner affective states, however contextually shaped and multiply realised, exhibit sufficient regularity at a basic categorical level to be inferred with high reliability. On this understanding, accepting Barrett's critique of basic emotion theory, for instance, is compatible with granting, for the sake of the present argument, that such systems could accurately detect inner affective states in a limited yet socially significant sense.

---

[7] I want to thank a reviewer for pointing this out to me.
[8] The use of discrete emotion-types should be understood as a pragmatic design choice rather than a substantive theoretical commitment. In practice, emotion-types are often favoured because they offer a tractable and communicable format for system outputs, particularly within explainability pipelines.

In theories of emotional expression, a typical requirement for an EE to count as one is that an EE, understood as a particular behaviour or physiological change, needs to be perceivable, that is, *humanly* perceivable. To better grasp the idea of "ideal ERTs", it seems that we need to slightly modify that requirement from humanly perceptibility to recognisability. By using the requirement of recognisability, I let go of the need for the relevant changes to be directly perceivable through ordinary visual or auditory perception. To qualify as recognisable, a pattern of (physiological and/or behavioural) changes needs only to exist, whether or not it has (yet) been detected or interpreted by human observers. Hence, this allows us to consider the possibility that human and technological recognition may rely on different *epistemic routes* to reach a given assessment of a particular inner affective state.

As already discussed above, human emotion *recognition* typically occurs *in situ* within shared contexts of meaning and expectation. It is inherently defeasible. Human perceivers interpret emotional expressions against a background of contextual information and can misread them, revise their interpretations, or seek clarification ("I thought she was angry but now she only seems tired"). Uptake is therefore both negotiated and constrained. Recognition here thus belongs to a communicative regime in which EEs typically function as essential contributions to an ongoing interaction, serving the emoter, the perceiver, and as a by-product, society as a whole (i.e. by coordinating expectations, managing relationships, and sustaining social norms).

Importantly, human recognition is limited both perceptually and normatively. Perceptually, humans have only partial access to the physiological and behavioural manifestations of emotional episodes (humans typically cannot track brain activity, heartbeat, or pupil dilation). Normatively, emotional interpretation remains constrained by role expectations, contextual relevance, and standards of appropriateness. This combination explains why ordinary recognition is generally reliable in everyday interaction, while also preserving space for ambiguity, discretion, and revision. The limits of human recognition are thus integral to the social-epistemic functions of intentional emotional expression, as they provide the necessary elements for them to do their work.

ERTs, by contrast, are designed to infer inner affective states directly. They abstract away from the intentional and communicative roles of emotional expressions and instead treat behavioural and physiological signals primarily as evidential inputs for inner state inference. What ERTs "recognise" are not expressions as expressions, but patterns of physiological and behavioural change taken to probabilistically indicate an underlying affective state (Scarantino 2017). From the perspective of a human perceiver, blushing may appear as the salient expression of embarrassment; from the perspective of an ideal ERT, this same episode may be embedded within a broader configuration of physiological and behavioural changes that humans cannot fully access. Ideal ERTs, as defined here, need not share the epistemic limits that structure human recognition. I am thus hypothesising here that ideal ERTs (in contrast to the simplified basic emotion theory on which many current ERTs rely) may discover and rely on complex patterns of physiological and behavioural changes, adapt their analysis to particular idiosyncrasies in individuals, and generate reliable outputs about inner affective states.[9]

This difference in epistemic orientation has important implications for affective communication. By privileging transparency over social negotiation, ideal ERTs risk reshaping what counts as sincerity. At first glance, ERTs' outputs may simply allow us to track more efficiently familiar manifestations of the sincerity condition in emotion theory, which often tie sincerity to whether an agent genuinely feels the emotion conventionally associated with a given expression, for instance, whether an apology is accompanied by genuine guilt. On this view, sincerity distinguishes expressions that reflect a "true" emotional state from those that merely simulate it.

However, while the sincerity condition is useful in, for instance, ontological work on emotions, ordinary moral and social practice suggests that a more nuanced understanding should be privileged (see Frankfurt 2005). Under ordinary conditions, sincerity in apology, reassurance, or gratitude is not primarily assessed by reference to *momentary affective alignment*, but by whether the expressed stance is genuinely endorsed and integrated into an agent's practical commitments over time. Residual resentment should not automatically render an apology insincere, nor should background anxiety make reassurance deceptive. Agents often express stances they recognise as normatively appropriate while still working through conflicting or unsettled affect.

---

[9] Note that, while this is a working hypothesis, it tracks recent empirical work in both neuroscience and affective computing, which keeps improving on the discovery of these hypothesised patterns that are not fully accessible to humans (Shu et al. 2018; Piretti et al. 2023; Yin et al. 2017).

In an ideal ERT environment, the sincerity condition risks being reinterpreted as a requirement of full affective transparency at the moment of expression. When sincerity is assessed primarily by reference to detected inner affective states, intentional emotional expressions lose their standing as authoritative acts through which agents endorse a stance toward others.

The space for committing to a position while still navigating emotional ambivalence is thereby narrowed. This threatens not only ordinary communicative practices, but also aspirational expressions that are central to moral and relational life.

Much moral and relational communication is aspirational. Agents express forgiveness they are still learning to inhabit, commit to patience they know will be tested, or signal trust they hope to build. Such expressions project a stance the agent is attempting to realise. They are epistemically valuable because they publicly initiate new normative expectations, both for oneself and for others, even when affect has not yet fully aligned. Aspirational (emotional) expressions rely on the defeasibility and interpretive openness of emotional communication. They presuppose that expressions may also be treated as commitments-in-progress rather than only as transparent reports of inner states.[10]

By providing granular snapshots of inner affective life, ideal ERTs risk defeating the possibility of these aspirational stances themselves. Indeed, in constricting emotional expressions to the role of a simple mirror of real-time affective profiles,[11] rather than acknowledging them as complex communicative acts with multiple functions, ideal ERTs undermine the broader expressive practices through which social meaning is negotiated. Once emotional communication is filtered through this diagnostic lens, the defeasible, context-sensitive interpretation that sustains the current broader range of expressive practices is systematically restricted.

The result is an epistemic environment in which individuals, but also social partners and even institutions, are encouraged to view emotional communication primarily through inner-state profiles. Intentional emotional expressions lose their standing as primary carriers of social meaning in certain contexts and become always secondary, expected to align with the affective state underlying them. This represents a significant technosocial risk. Ideal ERTs risk reorganising social life around a standard of affective transparency that is misaligned with the conditions under which expressive practices perform their social epistemic functions. Rather than supporting coordination, moral repair, interpersonal trust, and normative flexibility, emotional expressions risk being reduced to data points in a continuous affective audit, reshaping the very practices through which agents make sense of one another and hold one another to account.

# 4 EEs as generator of social goods and implication for ideal ERTs' regulation

Before turning to regulatory implications, it is important to acknowledge that affective transparency may, in some cases, plausibly prevent harm. Ideal ERTs could appear especially attractive in contexts where emotional expressions are used to mask malice, manipulation, or negligent judgement (i.e. consider lie detectors, border control, or law enforcement tools), for instance, when polite expressions conceal hostility or when displays of care are strategically deployed to exploit vulnerability. In such cases, rendering inner affective states legible might seem to protect those who would otherwise be exposed to deception or danger. Granting this point, however, does not undermine the present argument. The concern is not that affective transparency never yields local benefits, but that these local benefits are obtained at the price of other social goods that depend on partial opacity and epistemic friction within (affective) communication channels. Once this framework is normalised, the normative authority of intentional expressions is displaced by diagnostic assessment. Practices such as apology, reassurance, politeness, and restraint coexist with practices of deception, exploitation, or manipulation. Both depend on an interpretive environment in which expressions can function as context-sensitive, revisable acts addressed to others. While heightened scrutiny may be justified in narrowly specified, high-stakes contexts already governed by independent epistemic norms and institutional accountability, it is crucial to recognise that extending this logic to ordinary social interaction risks reorganising expressive life altogether. The resulting losses are a consequence of a structural transformation of the social-epistemic conditions under which intentional EEs currently operate.

If ideal ERTs reorganise the epistemic priorities of social life in ways that diminish the functions of intentional emotional expressions, then the harms that follow are not merely interpersonal. They affect the stability of social roles, undermine the capacity of individuals to act as normative agents, and reshape collective expectations about emotional conduct. Regulation must therefore take seriously the fact that intentional EEs are part of the normative infrastructure of social life. When this infrastructure is distorted, both social structures and individual agency are put at risk.

---

[10] For more general work on aspirations, see Callard (2018).

[11] This point is inspired by Vallor's 2024 book "The AI Mirror: How to Reclaim Our Humanity in an Age of Machine Thinking".

In many everyday, but also institutional settings, cooperation depends on shared expectations regarding how people express and receive emotional cues. These expectations are often tacit and role-dependent. Teachers are expected to project patience. Judges are expected to remain even-tempered. Colleagues are expected to use tact in disagreement. These expectations are not demands for perfect emotional transparency but demands for role-appropriate expressive conduct.

Ideal ERTs threaten these expectations by shifting attention away from the quality of the expressive act toward its correspondence with inner affect. When institutions begin to treat affective readouts as indicators of role fulfilment, the expressive behaviour becomes secondary. A teacher may be judged on whether they felt irritated rather than on how well they contained that irritation. A nurse may be evaluated on the presence of anxiety rather than the effectiveness of their reassurance. A political representative may be scrutinised for frustration during a debate rather than for the quality of their argument. These changes risk producing a social environment in which everyday interaction is shaped by continuous affective auditing.

This would be a form of affective determinism. Affective determinism, as used here, refers to a social-epistemic regime in which technology-based detection of affective states is treated as the authoritative indicator of an emoter's sincerity, trustworthiness, and moral standing.[12] In such an environment, emotional states come to be treated not as transient aspects of personal experience, but as direct predictors of moral standing, trustworthiness, or social value. The cumulative effect is a transformation of social norms in the direction of emotional transparency, despite the fact that many social goods *depend* on the preservation of expressive discretion.

One of the most significant implications of this analysis is that accuracy does not legitimise the deployment of ideal ERTs. Standard regulatory frameworks often treat accuracy as the central desideratum. The present chapter shows that this assumption fails to protect what is valuable in the case of emotional expressions (given "ideal" ERTs). The harms identified do not arise from inaccuracy but from the structural transformation of expressive practices that results from perfect accuracy.

A regulatory framework adequate for ideal ERTs must therefore depart from the accuracy paradigm and instead adopt what we may call a function-first perspective. The core question is not whether ERTs correctly identify affective states, but whether their use undermines the social functions that emotional expressions perform. This requires regulators to recognise intentional EEs as social goods that warrant protection. Such protection is a structural safeguard that allows social practices to function.

Below, I propose three normative (and interrelated) principles, that may be seen as proposals for the future development of regulatory guardrails against ERTs, which follow from the above analysis and may guide future regulation.

(1) Regulation should ensure that individuals retain reasonable control over how their emotional expressions are interpreted. This involves limiting the use of real-time affective analytics in interpersonal or role-dependent contexts.

(2) In contexts where emotional expressions perform relational functions, the relational meaning of expressive acts should be treated as primary. Legislators should treat the relational act (apology, reassurance, care, conflict de-escalation) as the relevant unit of analysis, not the affective state profile. This principle can justify restrictions on the admissibility or organisational use of emotional data.

(3) Regulation must recognise that many moral and emotional practices rely on aspirational expressions. Individuals should not be penalised or evaluated on the grounds that their expressed stance does not perfectly match their affective profile. This principle can support legal protections against institutional reliance on affective data in performance evaluations, disciplinary procedures, or other assessments of sincerity or trustworthiness.

Regarding the first principle, "reasonable control" should not be understood as individual consent over the collection of affective data, which is widely recognised as insufficient in institutional contexts characterised by power asymmetries. Rather, it concerns control over the interpretive authority of affective inferences. What requires regulation is not merely whether emotional data are captured, but how outputs are treated within decision-making structures. Importantly, the present analysis reveals that once affective outputs are institutionally treated as privileged indicators of sincerity, trustworthiness, or role fulfilment, the space for meaningful human oversight is already compromised. Existing regulatory frameworks, such as the European Commission's High-Level Expert Group on AI's Assessment List for Trustworthy AI (ALTAI), primarily operationalise control in terms of transparency, explainability, and the possibility of human-in-the-loop intervention at the level of individual decisions. In the case of ideal ERTs, the design and implementation levels are where regulation must intervene.

---

[12] Where genetic and neurobiological forms of determinism are typically relied on to explain behaviour by reference to underlying causal structures (Sapolsky 2023), here I rely on the concept of affective determinism to shed light on the reconfiguration of social evaluation itself, where the mechanism in operation treats detected affective states as authoritative indicators of moral character and convert communicative acts into real-time moral evidence: thereby bypassing the normative space in which agency, aspiration, and repair ordinarily operate.

Regarding the second principle, prioritising the relational meaning of EEs does not require denying the importance of safety, justice, or efficiency, but it does require recognising that some social practices are constituted by expressive interaction rather than merely supported by it. In many contexts of cooperation, deliberation, negotiation, or affiliation, intentional EEs are integral components of the practice itself. In such cases, allowing affective analytics to override relational meaning would not amount to a trade-off between competing values, but to a transformation of the practice in question. The appropriate regulatory standard is therefore practice-relative: where emotional expressions play a constitutive role, relational meaning sets a normative constraint on the admissibility of affective inference.

The third principle aims at protecting aspirational emotional expressions. This principle does not require institutions to distinguish genuine aspiration from, for instance, cynical performance (in Goffman's sense)[13] at the moment of expression. On the contrary, it requires resisting the temptation to make such distinctions on the basis of detected affective states altogether. Aspirational expressions are not vindicated through immediate affective alignment but through temporally extended patterns of commitment, responsiveness, and conduct. An apology, reassurance, or expression of patience may be aspirational precisely because affect has not yet fully aligned with the endorsed stance. Regulatory protection therefore consists in excluding affective analytics from evaluative contexts (such as performance assessment, disciplinary procedures, or judgements of sincerity) where agents are entitled to inhabit commitments-in-progress. The principal enforcement difficulty is the institutional appeal of affective data as a shortcut for moral judgement, which risks displacing longitudinal and practice-based assessments of agency with momentary affective snapshots.

Before closing, some remarks are in order. First, although the paper's illustrative cases focus on dyadic interactions such as apologiser-recipient, teacher-student, or parent-child relations to simplify the analysis, the regulatory implications of ERTs extend beyond these dual interactions to institutional and political contexts involving more than two individuals. Second, the proposed regulatory principles will vary significantly in their application depending on the social context of deployment. For instance, some systems directly mediate interpersonal interaction by shaping how emotional expressions are interpreted or responded to *in situ*, while others may be deployed institutionally for purposes of monitoring, evaluation, or governance. Individuals may also voluntarily use ideal ERTs for self-monitoring or self-optimisation purposes, raising distinct concerns. While the same underlying technology may be involved in each case, the relevant harms, power dynamics, and regulatory priorities differ across these contexts, and the regulatory principles proposed below should be understood as applying differentially to all ERT deployments.

To sum up, I have argued that ideal ERTs create technosocial risks that are structurally distinct from those typically addressed in existing regulatory frameworks. By reorganising epistemic priorities around affective transparency, they threaten the social goods generated by intentional emotional expressions and destabilise both social cohesion and individual agency. A regulatory framework that focuses exclusively on data protection, fairness, or accuracy will miss these deeper harms. Protecting the integrity of expressive practices therefore requires regulations that recognise expressive discretion as a social good and preserve the relational and aspirational dimensions of emotional expression.

If emotional expressions are part of the normative infrastructure that holds social life together, then the regulation of ideal ERTs is best understood as a matter of safeguarding the social conditions of human relationality.

# 5 Conclusion

The analysis presented in this chapter challenges the techno-optimist assumption that social life benefits from ever greater (affective) transparency. By examining the social functions of intentional emotional expressions through the lens of the prospect of ideal ERTs, the chapter has shown that the value of emotional communication lies not in its capacity to reveal inner affective states, but in its role within a complex ecology of coordination, repair, trust-building, and normative flexibility. Emotional expressions are not merely signals of internal feeling; they are enactments of stance and commitment that operate within shared social practices and depend on a background of partial opacity. This opacity is socially beneficial because it enables relational and moral agency.

---

[13] See Goffman's (1959) discussion on the Cynical Performer in contrast to the Sincere Performer.

Ideal ERTs risk jeopardising these conditions through their capacity to disclose inner affective states in real time. This exposure collapses epistemic friction, disrupts the relational meaning of expressive acts, and narrows the space for aspirational or forward-looking commitments. At a structural level, they shift interpretive priorities away from the socially recognised functions of emotional expression and toward an evaluative framework grounded in technology-based affective profiles. This shift risks producing forms of affective determinism in which emotional transparency becomes conflated with sincerity, trustworthiness, or moral standing.

The resulting harms are technosocial. ERTs risk undermining the normative infrastructure that makes coordinated action and social cohesion possible by distorting the very processes through which individuals navigate conflicting feelings, form commitments, and participate meaningfully in moral and social life.

Existing frameworks, which typically prioritise accuracy, fairness, and data protection, are not equipped to address the harms that follow from ideal ERTs. A future-ready regulatory approach must instead begin from a function-first perspective. It must recognise that intentional emotional expressions are social goods whose integrity is essential for the flourishing of individuals and for the stability of the social world. Safeguards must therefore protect expressive discretion and prohibit technological interference in the relational work of emotional expression. Real-time affective analytics should be restricted or banned in contexts where emotional communication performs a normative function, and individuals should retain the right to inhabit aspirational emotional stances without the continuous threat of interpretive collapse.

The ethical question raised in this chapter was thus whether a society that renders inner affective life fully legible can still sustain the normative practices that depend on controlled and context-sensitive forms of emotional expression. The analysis developed in this chapter supports a negative answer to this question. A society that renders inner affective life fully legible cannot sustain the normative practices that depend on controlled, defeasible, and context-sensitive forms of emotional expression. Moral repair, aspirational commitment, role-based trust, and social coordination all rely on expressive acts retaining normative authority that is not continuously overridden by (technology-based) authoritative interpretations of underlying affective states. By collapsing epistemic friction and reorienting interpretation towards technology-mediated affective profiles, ideal ERTs undermine the very conditions under which these practices generate social goods. If emotional expressions form part of the normative infrastructure of social life, then regulating ideal ERTs is not only a matter of optimising accuracy or transparency, but also of preserving the social and moral conditions that make human relationality, agency, and coordination possible.

**Declarations** This work was solely done by the author. The author has no relevant financial or non-financial interests to disclose.

**Availability of data and materials** N/A.

**Acknowledgements** This work has been supported by the SSHRC of Canada through doctoral funding (grant #752–2021-1281). I extend my gratitude to Dorota Mokrosinska and James McAllister for providing feedback on earlier drafts. I am also grateful to the team of the London AI and Humanity Project for given me the opportunity to do a research stay with them and to test early ideas. Finaly, many thanks to the reviewers for their extensive and invaluable feedback.

PreprintFischer AH, Sauter DA (2017) What the theory of affective pragmatics does and doesn't do. Psychological Inquiry, 28(2–3):190–193.

Frankfurt HG (2005) On Bullshit. Princeton University Press.

Goffman E (1959) The presentation of self in everyday life. Doubleday.

Green MS (2007) Self-expression. OUP.

Hareli S, Hess U (2010) What emotional reactions can tell us about the nature of others: An appraisal perspective on person perception. Cognition and emotion, 24(1):128-140.

Hareli S, Hess U (2012) The social signal value of emotions. Cognition & Emotion, 26(3):385-389.

Hess U, Hareli S (2015) The influence of context on emotion recognition in humans. In 11th IEEE international conference and workshops on automatic face and gesture recognition (FG), 3(1):1-6.

Hess U, Fischer AH (2013) Emotional mimicry as social regulation. Personality and Social Psychology Review, 17(2):142–157.

Hopster J (2021) What are socially disruptive technologies?. Technology in Society, 67, 101750.

Matsumoto D, Yoo SH, Fontaine J (2008) Mapping expressive differences around the world: The relationship between emotional display rules and individualism versus collectivism. Journal of cross-cultural psychology, 39(1):55-74.

Menges L, Weber-Guskar E (2025) Digital emotion detection, privacy, and the law. Philosophy & technology, 38(2), 77.

McStay A (2018) Emotional AI: The rise of empathic media. Sage.

Parkinson B, Fischer AH, Manstead ASR (2005) Emotion in social relations: Cultural, group, and interpersonal processes. Psychology Press.

Piretti L, Pappaianni E, Garbin C, Rumiati RI, Job R, et al. (2023) The neural signatures of shame, embarrassment, and guilt: A voxel-based meta-analysis on functional neuroimaging studies. Brain sciences, 13(4): 559.

Sapolsky RM (2023) Determined: A Science of Life without Free Will. Penguin Press.

Scarantino A (2017) How to do things with emotional expressions: The theory of affective pragmatics. Psychological Inquiry, 28(2-3):165-185.

Scarantino A (2019) Affective pragmatics extended: From natural to overt expressions of emotions. In Hess U, Hareli S (eds.) The social nature of emotion expression: What emotions can tell us about the world (pp.49-81). Springer.

Schwartz J (2020) Saying 'Thank You' and Meaning It. Australasian Journal of Philosophy, 98(4):718–731.

Searle JR (1979) Expression and Meaning: Studies in the Theory of Speech Acts. Cambridge University Press.

Shu L, Xie J, Yang M, Li Z, et al. (2018) A review of emotion recognition using physiological signals. *Sensors*, 18(7), 2074.

Smith N (2008) I Was Wrong: The Meanings of Apologies. Cambridge University Press.

Strawson PF (1962) Freedom and resentment. Proceedings of the British Academy.

Vallor S (2024) The AI Mirror: How to Reclaim Our Humanity in an Age of Machine Thinking. Oxford University Press.

van Kleef GA (2009) How emotions regulate social life: The Emotions as Social Information (EASI) model. Current Directions in Psychological Science, 18(3):184–188.

van Kleef GA (2016) The interpersonal dynamics of emotion: Toward an integrative theory of emotions as social information. Cambridge University Press.

Ventham E (2025) Beyond the Social Value of Privacy. Philosophy & Technology, 38(158).

Walker MU (2006) Moral Repair: Reconstructing Moral Relations after Wrongdoing. Cambridge University Press.

Wallace RJ (1994) Responsibility and the moral sentiments. Harvard University Press.

Wang Y, Song W, Tao W, Liotta A, Yang D, et al. (2022). A systematic review on affective computing: emotion models, databases, and recent advances. Information Fusion, 83:19-52.

Weiss R (1985) The morak and Social dimensions of gratitude. The Southern Journal of Philosophy, 23:491-501.

Yin Z, Zhao M, Wang Y, Yang J, Zhang J (2017) Recognition of emotions using multimodal physiological signals and an ensemble deep learning model. Computer methods and programs in biomedicine, 140:93-110.

Zhang J, Yin Z, Chen P, Nichele S (2020) Emotion recognition using multi-modal data and machine learning techniques: a tutorial and review. Information Fusion, 59:103-126.